%% file: thomson.tex
\documentclass[letterpaper,12pt]{article}

\usepackage[dvips]{graphicx}

\input{macros}

\begin{document}

\pagestyle{empty}
\vskip-10pt
\hfill {\tt hep-th/0210223}

\begin{center}
\vskip 2truecm
{\LARGE \bf  Thomson scattering of chiral tensors and \\[5mm] scalars
  against a self-dual string}
\vskip 2.5truecm

{\normalsize \bf P\"ar Arvidsson\footnote{\tt par@fy.chalmers.se}, Erik
  Flink\footnote{\tt erik.flink@fy.chalmers.se} and M{\aa}ns
  Henningson\footnote{\tt mans@fy.chalmers.se}}\\ 
\vskip 1truecm
{\it Department of Theoretical  Physics\\ Chalmers University of
  Technology and G\"oteborg University\\ S-412 96 G\"{o}teborg,
  Sweden}\\
\end{center}
\vskip 1cm
\noindent{\bf Abstract:}
We give a non-technical outline of a program to study the $(2, 0)$
theories in six space-time dimensions. Away from the origin of their
moduli space, these theories describe the interactions of tensor
multiplets and self-dual spinning strings. We argue that if the ratio
between the square of the energy of a process and the string tension
is taken to be small, it should be possible to study the dynamics of
such a system perturbatively in this parameter. As a first step in
this direction, we perform a classical computation of the
amplitude for scattering chiral tensor
and scalar fields (\ie the bosonic part of a tensor multiplet)
against a self-dual spinnless string.
\newpage
\pagestyle{plain}

\section{Introduction}
One of the most surprising discoveries in string theory during the
past decade is the existence of a new class of superconformal quantum
theories in six space-time dimensions, that do not contain dynamical
gravity. These so called $(2, 0)$ theories (the name refers to the
supersymmetry algebra under which they are invariant \cite{Nahm:1978}) made
their first appearance in the study of type~IIB string theory
compactified on a four-dimensional hyper-K\"ahler manifold
\cite{Witten:1995}. If we approach a point in the moduli space of the
hyper-K\"ahler manifold where it develops an isolated singularity, and
simultaneously take the string coupling constant to zero, a
six-dimensional theory at the locus of the singularity decouples from
the ten-dimensional bulk theory. The possible singularities of a
hyper-K\"ahler manifold obey an $ADE$-classification, \ie there are
two infinite series $A_r$, $r = 1, 2, \ldots$ and $D_r$, $r = 3, 4,
\ldots$, and three `exceptional' cases $E_6$, $E_7$ and $E_8$. Each
type of singularity gives rise to a different version of $(2, 0)$
theory, but these theories have no other discrete or continuous
parameters. However, the approach to the singular point in the moduli
space of the hyper-K\"ahler manifold and the possibility to turn on
holonomies of the two-form gauge potential of type IIB string theory
can be described by $5 r$ real moduli, where $r$ is the rank of the
singularity.

For the $A$-series of $(2, 0)$ theory, there is an alternative
description, in which the $A_r$ theory is regarded as the world-volume
theory on $r + 1$ parallel $M5$-branes in $M$-theory
\cite{Strominger:1996}. The $5 r$ moduli are then given by the relative
transverse positions of the $M5$-branes in the eleven-dimensional bulk
space. Similarly, one may describe the $D$-series of $(2, 0)$ theory
by including also a parallel orientifold plane in the
configuration. However, the $E_6$, $E_7$ and $E_8$ versions of $(2, 0)$ 
theory have no known realization within $M$-theory.

An intrinsically six-dimensional definition of the $(2, 0)$ theories,
which does not rely on their embedding into a ten- or
eleven-dimensional theory, is still lacking, though. In this paper, we
will suggest a program to study the $(2, 0)$ theories from a purely
six-dimensional perspective, with the hope that this work will
eventually lead to such a definition. Our approach is based on an
analogy between the $(2, 0)$ theories and non-abelian maximally
supersymmetric Yang-Mills theory in five space-time
dimensions. In the next section, we will give a rather non-technical
discussion of the degrees of freedom of the $(2, 0)$ theories and
their relationship to those of Yang-Mills theory. In section three, we
will then outline a perturbative scheme to study the quantum dynamics
of these degrees of freedom. We intend to systematically explore this
in forthcoming publications. As a first step, we will in the present
paper describe a purely classical computation of certain scattering
amplitudes. We will be working in a rather simplified toy model, which
is closely related to the $A_1$ version of $(2, 0)$ theory, but we
think that our analysis still captures many aspects of the complete
theory correctly.

A different approach to $(2,0)$ theory is described
in \cite{Dijkgraaf:2002}.

\section{The degrees of freedom of $(2, 0)$ theory}
The $M$-theory realization of the $A$-series of $(2, 0)$ theory
provides a convenient starting point for analyzing the degrees of
freedom of the theory. Starting with a single $M5$-brane, its
fluctuations may be described by a free massless $(2, 0)$ tensor
multiplet on the brane world-volume. Such a multiplet comprises a real
two-form gauge field with self-dual three-form field strength. We will
refer to such a field as a `chiral' two-form gauge field. It is a
singlet under the $SO (5) \simeq Sp (4)$ $R$-symmetry group which is
part of $(2, 0)$ supersymmetry. There are also five real scalar fields
in the vector representation of $SO (5)$, and four chiral spinor
fields in the vector representation of $Sp (4)$, that obey a
symplectic Majorana condition.  Upon quantization, these fields give
rise to massless particles, that may be described by their
transformation properties under the $SO (4)$ little group that leaves
the light-like six-momentum invariant. In this way, the two-form gauge
field provides three states in the rank two
anti-symmetric self-dual tensor representation of $SO (4)$, and the 
scalar fields provide five
$SO (4)$ invariant states. In total there are thus eight
bosonic states. Similarly, the spinor fields provide eight fermionic
states, that transform as two $SO (4)$ spinors. With $r + 1$ parallel
$M5$-branes, as is relevant for the $A_r$ version of $(2, 0)$ theory,
we have one such tensor multiplet for each $M5$-brane, but one linear
combination (namely the sum) decouples from the remaining theory and
will henceforth not be considered. (This is analogous to the fact that
the world-volume field theory of $r + 1$ parallel $D3$-branes in type
IIB string theory is a four-dimensional $N = 4$ super Yang-Mills
theory with gauge group $SU (r + 1)$ rather than $U (r + 1)$
\cite{Witten:1996}.)

Additional degrees of freedom come from open $M2$-branes that stretch
from one of the $r + 1$ $M5$-branes to another. From the
six-dimensional world-volume perspective, these are perceived as
strings, with a tension given by the tension of the $M2$-brane times
the transverse distance between the two $M5$-branes in question. At
the origin of the moduli space, the $M5$-branes are coincident and the
strings are thus tensionless. Interchanging the roles of the two
$M5$-branes amounts to changing the orientation of the string. The
string is self-dual in the sense that it has equal electric and
magnetic charges with respect to the chiral two-form gauge fields that
are part of the tensor multiplets on the two $M5$-branes. (These
charges obey Dirac quantization, and thus cannot be taken to be
small.) In the presence of such a string, the $(2, 0)$ supersymmetry
algebra develops a `central' charge, which is a vector under the $SO
(5, 1)$ Lorentz group and also a vector under the $SO (5)$
$R$-symmetry group \cite{Howe:1998}. The representation theory
of the $(2, 0)$ supersymmetry algebra with such a central charge
reveals that the string transforms non-trivially both under rotations
in the space transverse to the string and under the subgroup of the
$SO(5)$ $R$-symmetry group that is unbroken by the modulus
corresponding to the distance between the $M5$-branes. A minimal
`BPS-saturated' representation corresponds to a multiplet of straight
strings \cite{Gustavsson:2001sr}, while non BPS-saturated
representations describe waves propagating along the strings. These
degrees of freedom can be realized by complementing the scalar fields
on the string world-sheet, which represent the transverse space-time
coordinates of the string, with a set of fermionic fields.

We now turn to the realization of $(2, 0)$ theory within type IIB
string theory compactified on a hyper-K\"ahler manifold that develops
an $ADE$-singularity. Zero-modes of the massless type IIB fields will
then give rise to $r$ tensor multiplets, where $r$ is the rank of the
singularity. Furthermore, $D3$-branes that wrap around homologically
non-trivial two-spheres of the hyper-K\"ahler manifold will be
perceived as strings in six dimensions. There is in fact such a
two-sphere for each positive root of the corresponding $ADE$-type Lie
algebra (\ie $su (r + 1)$ for $A_r$, $so (2 r)$ for $D_r$, and the
$E_6$, $E_7$ and $E_8$ Lie algebras.) The tension of such a string is 
given by the tension of
the $D3$-brane times the area of the two-sphere. The latter is given
by the values of the moduli, and vanishes at the origin of the moduli
space, where the strings thus become tensionless. In summary, we have
been led to the following picture of the degrees of freedom of $(2,
0)$ theory: With each Cartan generator of the corresponding $ADE$-type
Lie algebra is associated a tensor multiplet of particles. With each
positive root generator of the Lie algebra is associated a multiplet
of straight strings. Finally, there are excitations that describe
propagating waves along these strings. 

In six dimensions, the degrees of freedom associated with the Cartan
generators and the root generators of the $ADE$-type Lie algebra are
rather different, and no Lie group symmetry is apparent in the
theory. However, if we compactify the theory on a circle, the tensor
multiplets associated with the Cartan generators give rise to massless
vector multiplets of maximally extended supersymmetry in five
dimensions. On the other hand, the string multiplets associated with
the root generators give rise to massive vector multiplets, if the
strings are straight and wound around the circle. The mass of these
multiplets is given by the tension of the string times the radius of
the circle. Strings that are not wound around the circle appear as
magnetically charged strings in five dimensions. We thus recover the
degrees of freedom of maximally supersymmetric Yang-Mills theory, with
an $ADE$-type gauge group spontaneously broken down to the maximal
torus group by the vacuum expectation values of the moduli fields.  As
we approach the origin of the moduli space, the massive vector
multiplets are un-Higgsed and also become massless vector
multiplets. The non-abelian gauge-symmetry of the theory then becomes
apparent. In six dimensions, this transition possibly corresponds to a 
theory
of tensionless strongly interacting self-dual strings, but we do not
yet have a clear picture of what this would mean. In particular, gauge
invariance plays a crucial role for the consistency of the
five-dimensional Yang-Mills theory, and should derive from some more
general principle in six dimensions, but exactly how this comes about
is still rather unclear. In this way, we may regard the compactified
$(2, 0)$ theory as the ultraviolet completion needed to make
five-dimensional maximally supersymmetric Yang-Mills theory into a
consistent theory.

\section{Investigating the quantum dynamics}
Their high degree of uniqueness is of course an attractive feature of
the $(2, 0)$ theories, but at the same time makes their study more
difficult. In particular, one might think that the absence of a small
parameter would make a perturbative analysis impossible, and force us
to define and solve the dynamics of the theory exactly. Fortunately,
this is not the case, as we will now explain: The approach that we
would like to advocate is to consider the theory at a point away from
the origin of the moduli space, so that the strings associated with
the root generators of the Lie algebra are tensile. (In any case, our
understanding of the degrees of freedom at the origin of the moduli
space is still too limited to allow us to work there.) Once the vacuum
expectation values of the moduli are fixed, a state in the theory can
be
characterized by giving the number of infinitely extended
approximately straight strings, together with their spatial directions
and momenta. In addition, the vibrational state of the string, which
describes waves propagating along the string, should be
specified. Finally, the state may contain a number of tensor multiplet
particle quanta of various types.

The energy of such a state is of course in general infinite, because
of the mass of the infinitely extended strings. However, since only
energy differences between different states affect the dynamics of the
theory, a more relevant statement is that the energy difference
between two states containing different numbers of strings is
infinite. This means that the Hilbert space of the theory decomposes
into separate superselection sectors, characterized by the number of
strings and their spatial directions. (This conclusion would of course
not hold if the theory is compactified on a circle, so that straight
strings have finite energy. Indeed, while charge is conserved in
Yang-Mills theory, particle number is not.) In each superselection
sector there is a ground state, containing only straight strings with
certain spatial directions and momenta. Above this ground state, there
are excited states that also contain quanta of propagating waves along
the strings and tensor multiplet particles. The energies of these
excitations only depend on the corresponding wavelengths and not on
the string tension. Furthermore, in the limit where the string tension
becomes large compared to the energies of these excitations, the
excitations decouple from each other. We are then left with a theory
of free tensor multiplets and free waves propagating on the strings,
\ie an infinite set of free harmonic oscillators. The latter theory is
of course exactly solvable, and can serve as a starting point for a
perturbative analysis. The parameter of such a perturbation expansion
is given by the square of the energy of a process (\eg the energy of
incoming tensor multiplet quanta), divided by the string tension.

So far, we have only discussed open infinitely extended strings. One
might also worry about the possible influence of closed
strings. However, on dimensional grounds, we would expect the
quantization of such strings to yield states with an energy of the
order of the square root of the string tension, so they can be safely
neglected in the limit that the tension goes to infinity. (If there
are any massless states in the spectrum of closed strings, other than
the tensor multiplets that we have been discussing, we expect them to
decouple from the rest of the theory.)

In view of the above considerations, it is natural to start exploring
$(2, 0)$ theory with a small number of infinitely extended
strings. Qualitatively new features then arise as this number is
increased. With no strings involved, we have the quantum theory of
free tensor multiplets, which is by now well understood. With a single
string, one could investigate the scattering of tensor multiplet
quanta against it. With two strings, the interaction of the strings
with each other can be studied. By analogy with Yang-Mills theory,
which in addition to the trilinear couplings contains also
quadrilinear couplings, one would expect that in addition to
interactions mediated by the exchange of tensor multiplet quanta,
there should also be direct string-string interactions. (On the other
hand, we do not expect any new couplings to appear as the number of
strings is increased further.) An additional subtlety is related to
the electric and magnetic charges of the strings. They imply that the
quantum wave function obeys Dirac quantization, \ie it is a section
of a line bundle rather than a function over the configuration space
of the strings. 

In a longer perspective, one would of course wish for a `second
quantized' formalism, capable of describing an arbitrary number of
strings. Presumably, this is best formulated in an expansion around
the origin of the moduli space, where the strings are tensionless. As
remarked above, such a formulation becomes necessary in order to
describe compactifications of $(2, 0)$ theory, in which straight
strings can have a finite mass and sectors with different number of
strings mix with eachother. As in many other examples in string and
$M$-theory, the dynamics of $(2, 0)$ theory is thus considerably
richer in a compactified space-time. 
 
\section{The classical limit}

In our future work, we intend to systematically explore the quantum
dynamics of $(2, 0)$ theory as outlined above, \ie to begin with we
will concentrate on the subsector of the theory that describes a
single string interacting with the quanta of a tensor multiplet. As
discussed in the previous section, this subsector should by itself
constitute a consistent theory. With only a single string involved, we
could without loss of generality restrict ourselves to the $A_1$
version of $(2, 0)$ theory, which only has a single type of string and
a single type of tensor multiplet. (This theory reduces to $SU (2)$
super Yang-Mills theory upon compactification.) We should then analyse
the quantum dynamics of this system perturbatively in a small
parameter given by the square of the energy of a process (\eg the
energy of incoming tensor multiplet quanta) divided by the string
tension. The first calculation to be performed is then to compute the
amplitude for scattering of tensor multiplet quanta off the string to
lowest order in the inverse string tension. However, before embarking
on this program of studying the quantum theory, we think that it is
worthwhile to ask if there are any questions that can be answered by a
purely classical computation. Because Dirac quantization effects
cannot be properly taken into account, we would be limited to the one
string subsector, and we would of course only trust the results to
lowest order in the energy square divided by the string tension.

A more serious objection to a classical computation is that we will
not be able to work in the complete $(2, 0)$ theory. The reason is
that the fermionic degrees of freedom of the tensor multiplet and on
the string world-sheet cannot be correctly represented in a classical
theory. We will therefore be restricted to a bosonic version of the
$A_1$ theory, which only contains a chiral two-form gauge field and
five scalar fields in the six-dimensional space-time, and the string
embedding coordinates on the world-sheet of the string. We do not
think that such a system could define a consistent quantum
theory. (Indeed, if such a theory existed, it would presumably give
rise to a hitherto unknown conformally invariant purely bosonic theory
upon compactification on a small two-torus.) Nevertheless, it is a
subtle and interesting classical mechanical system, which we think
deserves to be studied in its own right. The study of electrically
and/or magnetically charged objects of various dimensions interacting
with classical fields in a higher-dimensional space-time has a rather
long history, and many general results have appeared in the
literature. But, at least to our knowledge, the specific and rather
unique model that we are proposing has not been studied in detail. It
also contains many of the key ingredients of $(2, 0)$ theory, such as
chiral two-form gauge fields and self-dual strings, so we expect that
our results should be of some relevance for $(2, 0)$ theory. In a
certain sense, the calculations of the present paper are analogous to
the classical Thomson formula for the low-energy scattering of light
by an electrically charged spinnless point particle. This is
generalized by the Klein-Nishina formula, which is valid at arbitrary
energy and for a spinning particle. Summing over spin polarization and
going to the low-energy limit, the Klein-Nishina formula reproduces
the Thomson formula. Still, the Thomson and Klein-Nishina calculations
are rather different, and it is instructive to see how they yield the
same result in this limit. So, although we intend to describe a quantum
computation in the $(2, 0)$ theory in a forthcoming publication, we
think that the present classical computation is a logical first step
in exploring the dynamics of strings and particles in six dimensions.  

\subsection{The model}
Our fields are thus locally a chiral two-form $B$ 
and five scalar fields $\phi^A$, $A = 1, \ldots, 5$ 
in a  $(5+1)$-dimensional Minkowski space-time with signature
$(-+++++{}\!)$.
The values of the scalar fields
at spatial infinity $\phi^A_\infty$ constitute the moduli of the model. 
We may define these fields so that 
$\phi^2_\infty = \ldots = \phi^5_\infty = 0$. In fact, the fields
$\phi^2, \ldots, \phi^5$ then decouple completely from the rest of the
theory, and henceforth we will only consider a single scalar field
$\phi = \phi^1$. The string is described by its two-dimensional
world-sheet $\Sigma$, which we parametrize with some coordinates
$\tau$ and $\sigma$. Its embedding into space-time is given by a
set of functions $X^\mu (\tau, \sigma)$, $\mu = 0, 1, \ldots, 5$.

We wish to describe the dynamics of these degrees of freedom in a
Lagrangian formalism. As is well known, this poses a problem for the
chiral two-form $B$, which does not admit such a formulation
\cite{Witten:1997}. We therefore relax the condition that the three-form
field strength $H$ of $B$ be self-dual. However, our model will have 
the property that the `anti-chiral' part of $B$, \ie the part with
an anti self-dual field strength, is a `spectator field' that 
decouples completely from the other degrees of freedom. 
We then define the action of the model as
\beq
\eqnlab{action}
S = -\frac{1}{4 \pi} \int H \we *H -
\frac{1}{8 \pi} \int
d \phi \we * d \phi - \int_{\Sigma} B + \int_{\Sigma} \phi
\volsi.
\eeq
The first term is the generalized Maxwell action for the
two-form gauge-field $B$, while
the second is the kinetic term of the scalar field $\phi$. The third term
incorporates the electric coupling between the string and the gauge
field, and the fourth term is the Nambu-Goto term for the string. 
Here, \volsi{} is the volume form on the world-sheet \Si{} induced by its 
embedding into six-dimensional Minkowski space, \ie  
\beq
\volsi \equiv d \tau \we d \si \sqrt{-\mathrm{det} g_{\alpha \beta}} =
d \tau \we d \si \sqrt{-\mathrm{det} 
(\partial_\alpha X^\mu \partial_\beta X_\mu)} ,
\eeq
where $g_{\alpha \beta}$ is the induced metric on the world-sheet. The
indices $\alpha$ and $\beta$ take the values $0$ and $1$,
corresponding to $\tau$ and \si{}, respectively. The
string tension is thus formally given by the negative of the value of
the scalar field $\phi$ at the locus of the string world-sheet
$\Sigma$.

In addition to its electric coupling, the self-dual string should also 
have a magnetic coupling, which we put in `by hand'
\cite{Teitelboim:1986,Nepomechie:1985} by defining the field strength
$H$ as
\beq
H= dB + H_{\Si}.
\eeq
Here, $H_\Sigma$ is a three-form that only depends on the embedding
coordinates $X^\mu$ of $\Sigma$, and obeys the `Bianchi identity'
\beq
\eqnlab{bianchi}
d H_\Sigma = 2 \pi \desi .
\eeq
The four-form \desi{} is the `Poincar\'e dual' of the world-sheet $\Sigma$,
in the sense that 
\beq
\eqnlab{poincare}
\int_{\Si} s \equiv \int \desi \we s
\eeq
for an arbitrary two-form $s$ over Minkowski space. (Like in many other
formulas in this paper, we do not here notationally distinguish a form over
Minkowski space from its pullback to the world-sheet by the embedding
functions $X^\mu$.)

There are in fact no parameters that can be continuously adjusted in this 
action. 
The numerical constant preceding the first term of the action is fixed
by requiring that the anti self-dual part of $H$ decouples, 
while the constant in front of the second term is
determined by requiring that the net force between two infinitely long
and parallel straight strings vanishes, as we will see in the end of
this section. The constant multiplying the electric coupling term 
is fixed by demanding
invariance under `large' gauge transformations of the two-form gauge-field
$B$, whereas a constant multiplying the Nambu-Goto term can be absorbed
in a redefinition of the scalar field $\phi$. 

In the remainder of this section, we first derive the equations of motion
for the string and the fields. We then construct a configuration
describing a static, infinitely long, straight string. Finally, we
check that two parallel such strings do not exert any forces on
eachother.

\subsection{Equations of motion}
The equations of motion are obtained by requiring the action 
\eqnref{action} to be stationary under variations of
the generalized coordinates, \ie 
the gauge field $B$, the scalar field $\phi$, and the
coordinates $X^{\mu}$ that describe the embedding of the string 
world-sheet in Minkowski space. Because of the invariance of
the action under reparametrizations of the world-sheet, we may
impose two independent conditions on the $X^\mu$, though. A convenient
choice for our purposes is to parametrize the world-sheet by
$\tau = X^0 = x^0$ and $\sigma = X^5 = x^5$. The remaining $X^i$, 
$i = 1, \ldots, 4$, then describe the fluctuations around a static 
configuration containing an infinitely long, straight string in the 
$x^5$-direction. This parametrization may obviously break down for
large fluctuations, where the $X^i$ are no longer necessarily 
single-valued functions of $x^0$ and $x^5$.

We now start by varying the action with respect to $B$, and obtain the
equations of motion
\beq
\eqnlab{eomH}
d^* H = 2 \pi * \desi,
\eeq
which is consistent with the
Bianchi identity \eqnref{bianchi} and the self-duality condition
\beq
\eqnlab{selfduality}
H = * H ,
\eeq
as was required. This fixes the numerical coefficient of the first term
in \Eqnref{action}.

Next, we vary the action with respect to $\phi$, and obtain the equation
of motion
\beq
\eqnlab{eomphi}
d^* d \phi = - 4 \pi *(\desi \we \widehat{\mathrm{Vol}}_{\Si}),
\eeq
where $\widehat{\mathrm{Vol}}_{\Si}$ is some two-form over Minkowski
space whose pull-back to the world-sheet equals \volsi.

Finally, it remains to vary the action with respect to the string
coordinates $X^\mu$. This is a lot more 
involved than the previous calculations, therefore we will be more
explicit here. The reader may wish to skip this rather technical
derivation and jump directly to \Eqnref{eomx}.

To determine the variation of the first term in \Eqnref{action}, 
we first note that only $H_{\Si}$, and not $B$, depends on $X$. From the
Bianchi identity \eqnref{bianchi} follows that the variation of
$H_\Sigma$ is of the form
\beq
\delta H_\Sigma = 2 \pi \omega + d \Lambda ,
\eeq
where $\Lambda$ is some two-form, and the three-form $\omega$ is related
to the variation of the Poincar\'e dual $\delta_\Sigma$ as
\beq
\eqnlab{domega}
d \omega = \delta \delta_\Sigma.
\eeq
Thus, we get that
\beq
\eqnlab{deltaHH}
\de \left( - \frac{1}{4 \pi} \int H \we *H \right) = - \frac{1}{2 \pi}
\int H \we * \de H_{\Si} = - \int \om \we *H +
\int_{\Si} \La,
\eeq
where we also used the equations of motion \eqnref{eomH} for $H$. 
To proceed we
must determine the three-form $\om$ and the two-form \La. The
covariant expression for \desi{} is
\beq
\desi = - \int d \tau d \si \de^{(6)} (x - X(\tau, \si)) \Xdot_\mu
\Xprim_\nu * (\dxmn),
\eeq
which is easily verified by insertion in \Eqnref{poincare}. Here a dot
denotes a derivative with respect to $\tau$ while a prime
denotes a \si -derivative. From the variation of this quantity and
\Eqnref{domega}, we see that we can take 
\beq
\eqnlab{omega_cov}
\om = - \int d \tau d \si \de^{(6)} (x - X(\tau, \si)) \de X_{\mu} \Xdot_\nu
\Xprim_\rho * (\dxmnr),
\eeq
where we have used the contraction identities for
Levi-Civita symbols. 

To find \La, we need some knowledge
of the particular solution $H_{\Si}$ to \Eqnref{bianchi}. This,
however, is not
possible to obtain in a simple covariant form. With our choice of
parametrization of the world-sheet, as described in the beginning of this
subsection, we can write \desi{} as
\beqa
\desi & = & \frac{\eps_{ijkl}}{4!} \de^{(4)}(x-X) \Big( dx^i
\we dx^j \we dx^k \we dx^l - 4 \Xdot^i dx^0 \we dx^j \we dx^k \we dx^l
\nn \\ \eqnlab{desi} & & {} - 4 \Xprim^i dx^5 \we dx^j \we dx^k \we
dx^l + 12 \Xdot^i \Xprim^j dx^0 \we dx^5 \we dx^k \we dx^l \Big).
\eeqa
Here the dot and the prime denote $x^0$ and $x^5$ derivatives respectively,
and the four-index Levi-Civita symbol is defined by 
$\eps^{i_1 i_2 i_3 i_4} \equiv \eps^{0 i_1 i_2 i_3 i_4 5}$.
The first term describes a static, straight string while the following
terms allow for changes in time and for bending of the string. A solution
to the Bianchi identity \eqnref{bianchi} is then
\beqa
H_{\Si} & = & \frac{1}{3!} H^{\Si}_{ijk} \Big( dx^i \we dx^j \we dx^k -
3 \Xdot^i dx^0 \we dx^j \we dx^k - \nn \\
\eqnlab{Hnoncov1}
& & {} - 3 \Xprim^i dx^5 \we dx^j \we dx^k +
6 \Xdot^i \Xprim^j dx^0 \we dx^5 \we dx^k \Big),
\eeqa
where $H_{i j k}^\Sigma$ fulfills
\beq
\eqnlab{Hijk}
\pa_{[l} H^{\Si}_{ijk]} = \frac{\pi}{2} \eps_{lijk}
\de^{(4)}(x-X(\tau, \si)).
\eeq
This solution is of course unique only modulo the addition of a closed
three-form. Performing a variation of this $H_\Sigma$, we find that we can
take $\Lambda$ to be given by
\beq
\La = - \frac{1}{2} H^{\Si}_{ijk} \de X^i \Big( dx^j \we dx^k - 2
  \Xdot^j dx^0 \we dx^k - 2 \Xprim^j dx^5 \we dx^k + 2 \Xdot^j
  \Xprim^k dx^0 \we dx^5 \Big) .
\eeq
This calculation also yields the expression
\beqa
\omega & = & \frac{1}{6} \eps_{ijkl} \de^{(4)}(x-X) \de X^l \Big( dx^i
\we dx^j \we dx^k - 3 \Xdot^i dx^0 \we dx^j \we dx^k \nn \\
& & {} - 3 \Xprim^i dx^5 \we dx^j \we dx^k + 6 \Xdot^i \Xprim^j dx^0
\we dx^5 \we dx^k \Big),
\eeqa
in agreement with \Eqnref{omega_cov}. We can then write the two terms
on the right hand side of \Eqnref{deltaHH} in covariant form as
\beqa
- \int \om \we *H & = & - \int d^6 x
\de^{(4)}(x-X) (*H)_{\mu \nu \rho} \de X^{\mu} \Xdot^{\nu}
\Xprim^{\rho} \nn \\
& = & - \int_\Si d \tau d \si (*H)_{\mu \nu \rho} \de
X^{\mu} \Xdot^{\nu} \Xprim^{\rho}
\eeqa
and
\beq
\eqnlab{deltalambda}
\int_\Si \La = - \int_\Si d \tau d \si H^\Si_{\mu \nu \rho} \de
X^{\mu} \Xdot^{\nu} \Xprim^{\rho}.
\eeq

The variation of the electric coupling term is
\beq
\de \left( - \int_\Si B \right) = \de \left( - \int_\Si d \tau d \si
  B_{\mu \nu} \Xdot^{\mu} \Xprim^{\nu} \right),
\eeq
which after a partial integration becomes
\beq
\de \left( - \int_\Si B \right) =  - \int_\Si d \tau d \si (dB)_{\mu
  \nu \rho} \de X^{\mu} \Xdot^{\nu} \Xprim^{\rho}.
\eeq
We see that this term combines with the one obtained in
\Eqnref{deltalambda} into a single term involving $H=H_\Si + dB$. 
Thus, we have
\beq
\eqnlab{varHB}
\de \left( -\frac{1}{4 \pi} \int H \we *H - \int_\Si B \right) = - \int_\Si
d \tau d \si (H+ *H)_{\mu \nu \rho} \de X^{\mu} \Xdot^{\nu} \Xprim^{\rho},
\eeq
which shows the symmetry between the electric and magnetic
couplings explicitly. We see that an anti self-dual part of the field
strength $H$ would cancel from the variation.

Finally, the variation of the Nambu-Goto term is given by
\beq
\eqnlab{NGvar}
\delta \int d \tau d \sigma \phi \sqrt{ - \det g_{\alpha \beta}} 
= \int d \tau d \sigma \sqrt{ - \det g_{\alpha \beta}}
( \delta X^\mu \partial_\mu \phi + \phi \frac{1}{2} g^{\alpha \beta} 
\delta g_{\alpha \beta} ) .
\eeq
Here, $g^{\alpha \beta}$ is the inverse of the induced world-sheet metric
$g_{\alpha \beta}$. Inserting that
\beq
\delta g_{\alpha \beta} = \partial_\alpha \delta X^\mu \partial_\beta X_\mu
+ \partial_\alpha X^\mu \partial_\beta \delta X_\mu
\eeq
and performing a partial integration of the second term in
\Eqnref{NGvar}, we find that this equals
\beq
\int d \tau d \sigma \sqrt{ - \det g_{\alpha \beta}} \delta X^\mu
\left( \partial_\mu \phi - \partial_\nu \phi \partial^\alpha X_\mu
\partial_\alpha X^\nu - \phi D^\alpha D_\alpha X_\mu \right) ,
\eeq
where $D^\alpha$ is the covariant world-sheet derivative. One should note
that the $X^\mu$, while being a Minkowski space vector, are world-sheet
scalars. 

Altogether, we find that the equations of motion that follow from 
varying $X^\mu$ can be written as
\beq
\eqnlab{eomx}
\phi D^\alpha D_\alpha X_\mu = \partial_\mu \phi - \partial_\nu \phi
\partial^\alpha X_\mu \partial_\alpha X^\nu - (H + * H)_{\mu \nu \rho}
\Xdot^{\nu} \Xprim^{\rho},
\eeq 
evaluated at the locus of the string, \ie at $x^\mu = X^\mu$.

\subsection{Static solutions} \seclab{static}
We will now construct a configuration corresponding to 
an infinitely extended static straight
string along the $x^5$-direction located at $x^i = 0$, $i = 1, \ldots, 4$,
\ie with $X^i = 0$. For such a string, we have
\beq
\delta_\Sigma = \frac{\epsilon_{i j k l}}{4 !} \delta^{(4)} (x)
d x^i \wedge d x^j \wedge d x^k \wedge d x^l .
\eeq
The $H$ and $\phi$ fields are given by
\beq
H = \frac{1}{\pi} \frac{x_l}{|x|^4} \left(\frac{1}{3 !}
  \epsilon^{l}_{\ph{l} i j k} d x^i \wedge d x^j \wedge d x^k + d x^0
  \wedge d x^l \wedge d x^5 \right)
\eeq
and
\beq
\phi = \phi_\infty + \frac{1}{\pi} \frac{1}{|x|^2} .
\eeq
Here $|x| = \sqrt{x^i x_i}$, and $\phi_\infty$ is an arbitrary constant,
that gives the value of $\phi$ at spatial infinity.
It is straight-forward to verify that this configuration satisfies
the self-duality condition \eqnref{selfduality} and the equations
of motion \eqnref{eomH}, \eqnref{eomphi}, and \eqnref{eomx}.
(By rotational invariance in the transverse space, the right-hand side 
of the latter equation must vanish at $x^i = 0$.) 

Since the string is infinitely long, we expect that the energy of this
configuration diverges. It is however interesting to consider the energy
per unit length along the $x^5$-direction. To this end, we start by
determining the energy-momentum tensor for the fields
$\phi$ and $H$. It is most conveniently defined by coupling the theory
to some arbitrary six-dimensional metric $G_{\mu \nu}$ and computing
the variation of the action as this metric is varied. With
\beqa
S_0 & = & -\frac{1}{8 \pi} \int d^6 x \sqrt{- \det G} G^{\mu \nu} \pamd
\phi \pand \phi - \nn \\
& & {} - \frac{1}{4 \pi} \frac{1}{3!} \int d^6 x \sqrt{- \det G} 
G^{\mu \mu^\prime} G^{\nu \nu^\prime} G^{\rho \rho^\prime} 
H_{\mu \nu \rho} H_{\mu^\prime \nu^\prime \rho^\prime}
\eeqa
we get that
\beq
\de S_0 = \frac{1}{2} \int d^6 x \sqrt{- \det G} T^{\mu \nu} \de 
G_{\mu \nu} ,
\eeq
where the coefficient $T^{\mu \nu}$ is defined as the energy-momentum
tensor of the system. Computing $T^{\mu \nu}$ in this way for our
model, and then imposing the self-duality condition on $H$, we arrive at
\beq
\eqnlab{stress tensor}
T^{\mu \nu} = \frac{1}{4 \pi} \left( \pamu \phi \panu \phi -
  \frac{1}{2} G^{\mu \nu} \paru \phi \pard \phi \right) + 
\frac{1}{4 \pi} H^{\mu \rho \si} H^{\nu}_{\ph{\nu} \rho \si}.
\eeq
The energy per unit length of our configuration is now given by the 
integral of the $T^{0 0}$ component over the transverse space parametrized
by the $x^i$ plus the tension of string, \ie the negative of the value
of $\phi$ at
the locus of the string. Both of these contributions diverge, though.
We can regularize them by replacing the infinitesimally thin string at
$x^i = 0$ with a hollow cylinder of radius $\epsilon > 0$. This amounts to
integrating $T^{0 0}$ over the domain $|x| > \epsilon$ and evaluating
$\phi$ at $|x| = \epsilon$. One then finds that the divergences as
$\epsilon \rightarrow 0$ cancel, and we are left with an effective
string tension $T$, \ie energy per unit length, that equals
$-\phi_\infty$.

Finally, we point out that we may construct a static solution describing
parallel strings by a superposition of translations of the solution we
have just described. The equilibrium of such a configuration follows from
the fact that the right-hand side of \Eqnref{eomx}, \ie the force
exerted by the fields on a test string, vanishes identically everywhere
for the single string solution we have described.
Similar results are obtained in~\cite{Gustavsson:2001}.

\section{Scattering from an infinitely long string}
The analysis that we will perform is largely analogous to the
classical Thomson scattering, in which electromagnetic radiation is
scattered against a spinnless massive particle. (See \eg chapter 14
of \cite{Jackson:1999}.) This calculation is usually regarded to be valid
to lowest order in the electric charge of the particle, but to be
exact as a function of the mass of the particle. We will instead be
considering the scattering of a chiral two-form gauge field and scalar
field radiation against a self-dual string. Because of the
self-duality, the electric and magnetic charges of the string cannot
be taken to be small, so our calculation has to be exact in
them. Instead, we will take the square of the frequency of the
radiation to be small compared to the tension of the string, and work
to leading order in this ratio.

In this section, we will first describe the different kinds of plane
waves of radiation. We will then determine the response of a string to
such an incoming wave. The next step is to compute the reradiation
generated by the string motion. Finally, we analyze this outgoing
radiation far away from the string, where it appears as another plane
wave.

\subsection{Plane waves in six dimensions}

A plane wave is typically written as
$\mathrm{Re}(Ae^{i k_{\mu}x^{\mu}})$ and is specified by its (possibly
complex) amplitude $A$ and its wave vector $k_{\mu}$. The modulus of
the zeroth
component of the wave vector is the energy $E$ of the wave. Our fields
are massless and therefore $k_{\mu}$ is a light-like vector, thus
$k_{\mu} k^{\mu}=0$.

The $\phi$-field in our setup is a scalar field and therefore easily
described as (we omit the $\mathrm{Re}$-operators and keep in mind
that we should always take the real part of the final answer)
\beq
\phi = \phi_0 e^{i k_{\mu} x^{\mu}}.
\eeq
The $B$-field, however, is more complicated. Generally, it can be
written as
\beq
B_{\mu \nu} = b_{\mu \nu} e^{i k_{\rho} x^{\rho}},
\eeq
but there are certain conditions on the fifteen components of
$b_{\mu\nu}$. First, we choose a Lorentz-type gauge fixing condition
\beq
\eqnlab{gaugefix0}
\pamu B_{\mu \nu} = 0,
\eeq
which reduces to
\beq
\eqnlab{gaugefix1}
k^{\mu} b_{\mu \nu} = 0,
\eeq
for a plane wave. However, this condition does not fix the gauge
entirely, since we may perform a transformation
\beq
b_{\mu \nu} \rightarrow b_{\mu \nu} + k_{\mu} \La_{\nu} - k_{\nu}
\La_{\mu}
\eeq
without altering $k^{\mu} b_{\mu \nu}$, if $k^{\mu} \La_{\mu}=0$. We
also see that $\La_{\mu}=k_{\mu}$ is a trivial transformation that
does not alter $b_{\mu \nu}$. These observations reduce the number of
independent components in $\La_{\mu}$ from six to four, which means
that we need four additional conditions to fix the gauge entirely. We
may take these to be
\beq
\eqnlab{b0a=0}
b_{0a}=0,
\eeq
where $a$ ranges over the spatial indices, \ie from 1 to 5. (This
yields four independent conditions since $k^a b_{a0}=0$ from
\Eqnref{gaugefix1}.)

This gauge fixing leaves us with six independent components in $b_{\mu
  \nu}$. We must also impose a self-duality condition which reduces
  this number to three, which are interpreted as the
  polarizations of the gauge field. The self-duality condition reads
\beq
H_{\mu \nu \rho} = \frac{1}{3!} \eps_{\mu \nu \rho}^{\ph{\mu \nu \rho}
  \si \ka \la} H_{\si \ka \la}
\eeq
in terms of the field strength $H$. For a plane wave $B$, such that
  $H=dB$, this becomes
\beq
k_{\mu} B_{\nu \rho} + k_{\nu} B_{\rho \mu} + k_{\rho} B_{\mu \nu} =
\frac{1}{2} \eps_{\mu \nu \rho}^{\ph{\mu \nu \rho} \si \ka \la}
k_{\si} B_{\ka \la},
\eeq
which immediately gives that
\beq
\eqnlab{selfdualB}
B_{ab} = \frac{1}{2} \eps_{ab}^{\ph{ab}cde} B_{cd} \hat{k}_e,
\eeq
where the hat on $k_e$ indicates that it is normalized to have unit
modulus. This may be interpreted by saying that $B$ is self-dual in
  the four-dimensional space orthogonal to $k^a$.

Now, consider the energy-momentum tensor for these
plane waves. By applying \Eqnref{stress tensor} we get that
\beq
T^{\mu \nu} = \frac{1}{4 \pi} \Big[ \mathrm{Re} \left( \phi_0 i
  k^{\mu} e^{i k_{\la}x^{\la}} \right) \mathrm{Re} \left( \phi_0 i
  k^{\nu} e^{i k_{\tau} x^{\tau}} \right) + \mathrm{Re} \left(i
  k^{\mu} b_{\rho \si} e^{i k_{\la}x^{\la}} \right) \mathrm{Re}
\left(i k^{\nu} b^{\rho \si} e^{i k_{\tau} x^{\tau}} \right) \Big]
\eeq
where we have used that $k^{\mu} k_{\mu}=0$.
Note the correspondence between $b_{\rho \si}b^{\rho \si}$ and
$\phi_0^2$ in this expression, meaning that we have the same
normalization with respect to energy for the two types of waves.

\subsection{String vibrations}

Now that we have explored the area of plane waves, we turn to
the actual scattering situation. The philosophy goes as follows: We
start from the static case considered previously but add an incident
plane wave. This wave
will cause the string to vibrate, but the deviations from the static
case will be small if the string tension $-\phiinf$ is large. The
string vibrations will in turn cause emission
of new waves, which however will be much smaller in amplitude than the
incident plane wave. This means that we may neglect back reaction, \ie
the action of the outgoing waves on the string. 

The general equations of motion for the fields $H(x)$ and $\phi(x)$
and for $X(x^0,x^5)$ are given by \Eqsref{eomH}, \eqnref{eomphi} and
\eqnref{eomx}. They are
\beqa
dH & = & 2 \pi \desi \\
d^*d \phi & = & -4 \pi \Omsi \\
\eqnlab{eomX2}
\phi (\Xddot_i-\Xdprim_i) & = & - \pa_i \phi + 2 (H_{i05} +
H_{ij5}\Xdot^j + H_{i0j} \Xprim^j) + \ordo(X^2)
\eeqa
where we have used that $H=*H$ and introduced $\Omsi \equiv *(\desi
\we \widehat{\mathrm{Vol}}_{\Si})$. Anticipating that the string
fluctuations $X^i$ will be small, we have only expanded the last
equation to second order in these.

Since the outgoing waves are suppressed by a factor
$\frac{1}{\phiinf}$, which is small when the string tension approaches
infinity, we may expand these equations in powers of \vepphi, where
\veps{} is some energy scale introduced to get a dimensionless
expansion parameter. This gives
\beqa
H & = & \Hst + \Hin + \vepphi \Hout +
\ordo\left(\vepphis\right) \\
\phi & = & \phi_{\infty} +\phist + \phiin + \vepphi \phiout
+ \ordo\left( \vepphis \right) \\
X^i & = & 0 + \vepphi \Xth^i +
\ordo\left(\vepphis\right) \\
\desi & = & \dest + \vepphi \deth +
\ordo\left(\vepphis\right) \\
\Omsi & = & \Omst + \vepphi \Omth +
\ordo\left(\vepphis\right).
\eeqa
Here ``st'' refers to the static configuration described in
subsection~\secref{static}. (\phist{} stands for the static
$\phi$-field minus the constant \phiinf.) \Hin{} and \phiin{} denote the
incoming plane waves. Quantities denoted ``Th'' (for Thomson) describe
the response of the string, whereas reradiation is described by
\Hout{} and \phiout. Note that we have explicitly extracted the
appropriate powers of \vepphi.

This allows us to rewrite the equations of motion order by order in
\vepphi. Observing that the static fields must obey the
equations of motion for a static, straight string, namely
\beqa
d \Hst & = & 2 \pi \dest \\
d^*d \phist & = & -4 \pi \Omst \\
\pa^i \phist + 2 \Hst^{i05} & = & 0,
\eeqa
and that $\Hst^{ij5}=\Hst^{i0j}=0$, we get the following equations:
\beqa
d \Hin & = & 0 \eqnlab{Hineom} \\
d^* d \phiin & = & 0 \\
d \Hout & = & 2 \pi \deth \eqnlab{Houteom} \\
d^*d \phiout & = & - 4 \pi \Omth \eqnlab{phiouteom} \\
\veps^2(\Xddotth^i - \Xdprimth^i) & = & - \pa^i \phiin - 2
\Hin^{i05}. \eqnlab{Xeom}
\eeqa
We see that the outgoing waves do not appear in the equation for
$\Xth^i$, which allows us to solve for the string vibrations directly,
as was anticipated. To solve \Eqnref{Hineom}, we introduce a gauge
field \Bin{} such that $\Hin=d \Bin$. We take this to be a plane
wave, satisfying the gauge fixing conditions \eqnref{gaugefix0} and
\eqnref{b0a=0}.

The next step is to solve \Eqnref{Xeom} for $\Xth^i$. We start with the
simplest case where $\Hin=0$ and $\phiin=\phi_0 e^{i k_{\mu}
  x^{\mu}}$ with the wave vector $k_{\mu}=-E(1,0,0,0,\sin \theta,\cos
\theta)$. The situation is depicted in \Figref{scattin}.
\begin{figure}[!htb]
\begin{center}
\includegraphics[width=10cm]{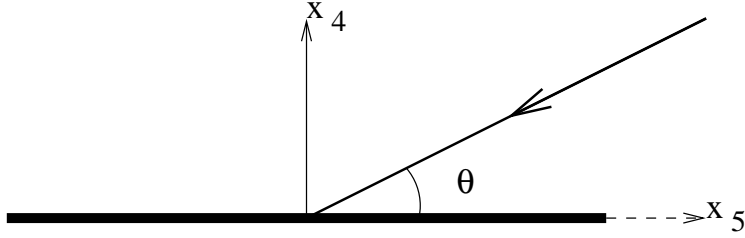}
\caption{Plane wave incident on an infinitely long string in the
  $x^5$-direction.}
\figlab{scattin}
\end{center}
\end{figure}

For this incident plane wave, we get that
\beq
\pa^i \phiin = -iE\sin \theta \de^i_{\ph{i}4} \phi_0 e^{-iE(x^0+x^4 \sin
  \theta + x^5 \cos \theta)},
\eeq
which when inserted in \Eqnref{Xeom} yields
\beq
\veps^2(\Xddotth^i - \Xdprimth^i) = iE \sin \theta \de^i_{\ph{i}4} \phi_0
e^{-iE(x^0 + x^5 \cos \theta)},
\eeq
since we may, to first order, let $X_i=0$ in the exponent. Remember
that \Eqnref{Xeom} should be evaluated on the world-sheet, which in
our parametrization is equivalent to letting $x^i=X^i$.

This equation is solved by making the ansatz
\beq
\eqnlab{ansatz}
\Xth^i = \xi^i e^{-iE(x^0 + x^5 \cos \theta)},
\eeq
which immediately gives that
\beqa
\eqnlab{xi_phi}
\Xth^i & = & - \frac{i \phi_0}{\veps^2 E \sin \theta}
\de^{i}_{\ph{i}4} e^{-iE(x^0 + x^5 \cos
  \theta)}.
\eeqa
Thus, an incident $\phi$-field in the $x^4-x^5$ plane causes string
vibrations only in the $x^4$-direction (to first order).

Let us next consider the case when $\phiin=0$ and $\Binu_{\mu \nu}=b_{\mu
  \nu} e^{i k_{\rho}x^{\rho}}$ with the same $k_{\mu}$ as in the
  previous case. This means that the field strength $\Hin$ becomes:
\beq
\Hinu_{\mu \nu \rho} = i(k_{\mu} b_{\nu \rho} + k_{\nu} b_{\rho \mu} +
  k_{\rho} b_{\mu \nu}) e^{i k_{\sigma}x^{\sigma}},
\eeq
which when inserted into \Eqnref{Xeom} yields
\beq
\veps^2(\Xddotthu_i - \Xdprimthu_i) = 2i(k_i b_{05} + k_0 b_{5i} + k_5
  b_{i0})e^{-iE(x^0 + x^5 \cos \theta)}.
\eeq
Using the ansatz \eqnref{ansatz} and the condition \eqnref{b0a=0} we
  get
\beq
\eqnlab{xi_B}
\Xthu_i = - \frac{2i}{\veps^2 E \sin^2 \theta} b_{i5} e^{-iE(x^0 +
  x^5\cos \theta)}.
\eeq
Since we are treating the model in a linearized approximation, it is
  of course possible to consider a general superposition of several
  kinds of incoming plane waves.

\subsection{Outgoing radiation}

The next step in our solution of the scattering problem is to solve
\Eqsref{Houteom} and \eqnref{phiouteom}. To do this, we need to find
expressions for $\deth$ and $\Omth$. These are found by expanding
\Eqnref{desi} in orders of \vepphi{} by replacing $X$
with $\vepphi \Xth$. Doing this, we see that
\beqa
\deth & = & -\frac{1}{4!} \eps_{ijkl} \Big[\Xth^{i_1} \pa_{i_1}
\de^{(4)}(x) dx^i \we dx^j \we dx^k \we dx^l + {} \nn \\
& & {} + \de^{(4)}(x) \Big(4
\Xdotth^i dx^0 \we dx^j \we dx^k \we dx^l + 4 \Xprimth^i dx^5 \we dx^j
\we dx^k \we dx^l\Big)\Big] \quad \\
\Omth & = & -\Xth^i \pa_i \de^{(4)}(x).
\eeqa 
\Eqnref{Houteom} is hard to solve as it stands without breaking the
self-duality condition, but we may decompose
the self-dual \Hout{} as
\beq
\Hout = \Hel + * \Hel
\eeq
where \Hel{} satisfies
\beqa
d \Hel & = & 0 \\
d^* \Hel & = & 2 \pi * \deth.
\eeqa
This means that we can find a \Bel{} such that $\Hel=d \Bel$. \Bel{}
should satisfy the same gauge fixing condition \eqnref{gaugefix0} as
\Bout. In this gauge, the equations of motion for \phiout{} and \Bel{}
are
\beqa
\pamd \pamu \phiout & = & 4 \pi \Xth^i \pa_i \de^{(4)}(x) \\
\pamd \pamu \Belu_{ij} & = & 0 \\
\pamd \pamu \Belu_{i5} & = & 2 \pi \Xdotthu_i \de^{(4)}(x) \\
\pamd \pamu \Belu_{0i} & = & - 2 \pi \Xprimthu_i \de^{(4)}(x) \\
\pamd \pamu \Belu_{05} & = & 2 \pi \Xth^i \pa_i \de^{(4)}(x)
\eeqa
To solve these, we need to find the Green function satisfying
\beq
\pamd \pamu D(t,\bx;t',\bx') = \de(t-t')
\de^{(4)}(x-x')\de(x^5 - x'^5),
\eeq
where obviously $t\equiv x^0$ and $\bx=(x^1,x^2,x^3,x^4,x^5)$. The
retarded solution to this equation is (\cf \cite{Jackson:1999},
Ch. 12)
\beqa
D_{\mathrm{ret}}(t,\bx;t',\bx') & = & - \frac{1}{8 \pi^2}
\theta(t-t') \Big[ \frac{1}{|\bx-\bx'|^3} \de(t-t'-|\bx-\bx'|) + {} \nn \\
& & {} + \frac{1}{|\bx-\bx'|^2} \de'(t-t'-|\bx-\bx'|) \Big],
\eeqa
where $\theta$ is the ordinary Heaviside step function.

We need to calculate the convolutions of this Green function with both
$\de^{(4)}(x)$ and $\pa_i \de^{(4)}(x)$ multiplied by the exponential
factor in the expression for $\Xth$ (it is the only part of $\Xth$
with dependence on $x$). The integrals thus obtained are not
analytically solvable, but we may solve them in the case when
$r\equiv \sqrt{x^i x_i}$ is very large compared to the wavelength. This is
equivalent to saying that $rE \gg 1$, \ie we look in the far-field
region.

We have that
\beqa
f(t,\bx) & = & \int dt' d^5x' D_{\mathrm{ret}}(t,\bx;t',\bx')
\de^{(4)}(x') e^{-iE(x'^0+x'^5\cos \theta)} \\
& = & - \frac{1}{8 \pi^2} \int dx'^5 \Big[ \frac{1}{(r^2 +
  (x^5-x'^5)^2)^{3/2}} - {} \nn \\
\eqnlab{f_int}
& & {} - \frac{iE}{r^2 + (x^5-x'^5)^2} \Big] e^{-iE(t-(r^2 +
  (x^5-x'^5)^2)^{1/2} + x'^5 \cos \theta)},
\eeqa
which is as far as we get without doing approximations. The geometry
of the present situation is  shown in \Figref{scattout}. 
\begin{figure}[!htb]
\begin{center}
\includegraphics[width=10cm]{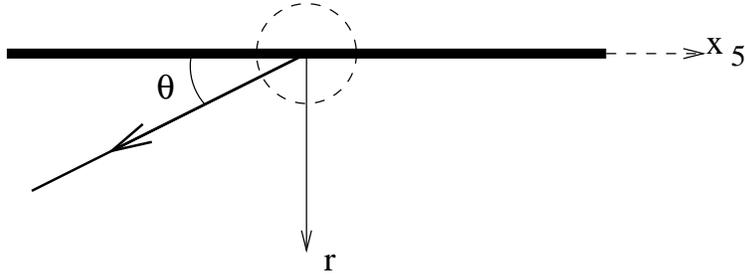}
\caption{Outgoing wave in the $r$-$x^5$ plane.}
\figlab{scattout}
\end{center}
\end{figure}
The angle $\theta$ in this figure must be the same as the one in
\Figref{scattin} since both $k_5$ and $k_{\mu} k^{\mu}$ are unaffected
by the scattering. Note though that the plane in \Figref{scattout}
does not have to be $x^4-x^5$, the vector $r$ may in fact point in any
direction perpendicular to
the string. This means that the different possible outgoing wave
vectors span a three-sphere. The dominating contribution to the field
at a point $x^{\mu}$ is from the part of the string inside the dashed
circle. This means that if we write
\beq
x'^5 = x^5 + \frac{r}{\tan \theta} + s \sqrt{\frac{r}{E}},
\eeq
then only small $s$-values should contribute significantly. Performing
this substitution, we may expand the exponential and the denominators
in the integral \eqnref{f_int}. The result of this substitution and
the subsequent Taylor expansions is
\beq
f(t,\bx) = - \frac{1}{8 \pi^2} e^{-iE(x^0 - r \sin \theta +
  x^5 \cos \theta)} \int ds \sqrt{\frac{r}{E}} \frac{-iE}{r^2} \sin^2
\theta e^{\frac{i}{2}\sin^3\theta s^2}  
\Big[ 1 + \ordo \Big(\frac{1}{\sqrt{rE}} \Big) \Big]
\eeq
which becomes, after performing the Gaussian integral,
\beq
f(t,\bx) = \sqrt{\frac{E \sin \theta}{32 \pi^3
    r^3}} i e^{i \frac{\pi}{4}} e^{-iE(x^0 - r \sin \theta +
  x^5 \cos \theta)} \Big[ 1 + \ordo \Big(\frac{1}{\sqrt{rE}} \Big) \Big].
\eeq

In the same way, we may calculate
\beq
g_i(t,\bx) = \int dt' d^5x' D_{\mathrm{ret}}(t,\bx;t',\bx')
\pa'_i \de^{(4)}(x') e^{-iE(x'^0+ x'^5 \cos \theta)},
\eeq
which yields
\beq
g_i(t,\bx) = -\frac{x_i}{r} \sqrt{\frac{E^3 \sin^3 \theta}{32 \pi^3
    r^3}} e^{i \frac{\pi}{4}} e^{-iE(x^0 - r \sin
  \theta + x^5 \cos \theta)} 
\Big[ 1 + \ordo \Big(\frac{1}{\sqrt{rE}} \Big) \Big].
\eeq

If we use the results just obtained together with the equations of
motion for \phiout{} and \Bel{} we get, to lowest order, the outgoing
waves (we are only interested in the particular solutions, not the
homogeneous)
\beqa
\eqnlab{phiout}
\phiout & = & 4 \pi \xi^i g_i \\
\Belu_{ij} & = & 0 \\
\Belu_{i5} & = & -2 \pi i E \xi_i f \\
\Belu_{0i} & = & 2 \pi i E \cos \theta \xi_i f  \\
\Belu_{05} & = & 2 \pi \xi^i g_i,
\eeqa
where $\xi^i$ is defined in \Eqnref{ansatz}. Having found \Bel{} we
can easily find $\Hout=d \Bel + * d \Bel$. Defining the amplitude
$h'_{\mu \nu \rho}$ by
\beq
\vepphi \Houtu_{\mu \nu \rho} \equiv h'_{\mu \nu \rho} e^{-iE(x^0
  - r \sin \theta + x^5 \cos \theta)},
\eeq
we get, after a small calculation, that
\beqa
h'_{ijk} & = & \vepphi \sqrt{\frac{ E^5 \sin^3
    \theta}{8\pi r^3}} i e^{i \frac{\pi}{4}} \sin \theta \eps_{ijk i_1}
\left( \xi^{i_1} - \frac{x^{i_1} x^{i_2} \xi_{i_2}}{r^2}\right) \\
h'_{0ij} & = & \vepphi \sqrt{\frac{ E^5 \sin^3
    \theta}{8\pi r^3}} i e^{i \frac{\pi}{4}} \left(\cos \theta \frac{x_i
    \xi_j - x_j \xi_i}{r} - \eps_{ij}^{\ph{ij}kl} \frac{x_k \xi_l}{r}
\right) \\
h'_{5ij} & = & \vepphi \sqrt{\frac{ E^5 \sin^3
    \theta}{8\pi r^3}} i e^{i \frac{\pi}{4}} \left( \frac{x_i
    \xi_j - x_j \xi_i}{r} - \cos \theta \eps_{ij}^{\ph{ij}kl}
  \frac{x_k \xi_l}{r} \right) \\
h'_{05i} & = & \vepphi \sqrt{\frac{ E^5 \sin^3
    \theta}{8\pi r^3}} i e^{i \frac{\pi}{4}} \sin \theta \left(\xi_i -
  \frac{x_i x_{i_1}\xi^{i_1}}{r^2} \right).
\eeqa
Analogously, we define the amplitude $\phi'_0$ by
\beq
\vepphi \phiout = \phi'_0 e^{-iE(x^0 - r \sin \theta + x^5 \cos
  \theta)}.
\eeq
\Eqnref{phiout} then yields that
\beq
\eqnlab{phi'0}
\phi'_0 = - \vepphi \sqrt{\frac{E^3 \sin^3 \theta}{2 \pi r^3}} \frac{x_i
  \xi^i}{r} e^{i \frac{\pi}{4}}.
\eeq
We see that the fields decrease as $r^{-3/2}$ (at least, depending on
direction). This makes sense, since it means that the energy decreases
as $r^{-3}$, which is consistent with the fact that the possible
outgoing wave vectors span a three-sphere. The wave vector $k'_{\mu}$
of the outgoing wave is
\beqa
k'_0 & = & -E \\
k'_i & = & E \sin \theta \frac{x_i}{r} \\
k'_5 & = & -E \cos \theta,
\eeqa
meaning that the wave vector indeed points as in \Figref{scattout}.

Far away from the string, \Hout{} can be viewed as a plane wave, the
only important $x$-dependence is then in the exponent. The amplitude
factor may be taken to be constant. This means that $d \Hout = 0$;
thus we can find a \Bout{} such that $\Hout = d \Bout$. If we write
\beq
\eqnlab{bdef}
\vepphi\Boutu_{\mu\nu} = b'_{\mu \nu} e^{-iE(x^0 - r \sin \theta + x^5
  \cos \theta)},
\eeq
the apparent $b'_{\mu \nu}$ that yields the correct \Hout{} is then
given by
\beqa
b'_{ij} & = & \vepphi\sqrt{\frac{E^3 \sin^3 \theta}{8 \pi r^3}} e^{i
   \frac{\pi}{4}} \left( \eps_{ijkl} \frac{x^k \xi^l}{r} - \frac{x_i
     \xi_j - x_j \xi_i}{r} \cos \theta \right) \\
b'_{0i} & = & 0 \\
\eqnlab{b'i5}
b'_{i5} & = & \vepphi\sqrt{\frac{E^3 \sin^3 \theta}{8 \pi r^3}} e^{i
   \frac{\pi}{4}} \sin \theta \left(\xi_i - \frac{x_i
     x_j \xi^j}{r^2} \right) \\
b'_{05} & = & 0.
\eeqa
It is also clear that this \Bout{}
satisfies the gauge fixing conditions \eqnref{gaugefix1} and
\eqnref{b0a=0} \emph{if} considered as a plane wave.

\subsection{Presenting the results}

The results of the scattering problem may be presented in a more
transparent way by exploiting the symmetry of the problem. When
choosing the plane spanned by the incident wave vector and the string,
the original $SO(5)$ symmetry is broken down to $SO(3)$. It is
therefore convenient to introduce a three-dimensional vector space
spanned by
\beq
\eqnlab{b_a def}
b_a \equiv \frac{2}{\sin^2 \theta} b_{ab} n^b,
\eeq
where $n$ is a unit vector pointing in the string
direction and the prefactor is determined by requiring that
\beq
b^a b_a = b^{ab} b_{ab}.
\eeq
The angle $\theta$ is still taken to be the angle between the incident
wave and the string, which in terms of $n$ and $k$ is given by
\beq
\cos \theta = - n^a \kh_a,
\eeq
where the hat denotes a unit vector.
Analogously, we define the angle $\varphi$ as
\beq
\cos \varphi = \kh^a \kh'_a.
\eeq
Obviously, the vector $b$ is orthogonal to $n$ and $k$, hence it has
only three independent components (polarizations).

The inverse to \Eqnref{b_a def} is given by
\beq
b_{ab} = b_{[a}n_{b]} + \cos \theta b_{[a} \kh_{b]} +
  \frac{1}{2} \eps_{ab}^{\ph{ab}cde}b_c n_d \kh_e.
\eeq

In our choice of coordinates $n^a=(0,0,0,0,1)$, meaning that
$b_5=0$. Thus we may omit this component and work only with
\beq
b_i=\frac{2}{\sin^2 \theta} b_{i5}.
\eeq

After these preliminaries, we may now express the scattered fields in
terms of the incident in the following way:
\beqa
\eqnlab{aij,ci}
b'_i & = & A_{ij} b^j + C_i \phi_0 \\
\eqnlab{ej,f}
\phi'_0 & = & E_j b^j + F \phi_0,
\eeqa
where $b'_i$ is defined from $b'_{ab}$ in the same way as $b_i$ is
defined from $b_{ab}$. The plus-sign in these relations should be
interpreted with caution. The contributions from $B$ and $\phi$ can be
added like this \emph{only} if they have the same wave vector $k$ (and
therefore the same energy). If this is not the case, they can still be
added (since the problem is linear) but one must take into account
that $\phi$ and $B$ have different wave vectors.

We will now determine the coefficients $A$, $C$, $E$ and $F$ in these
relations, which is done by using \Eqsref{xi_phi},
\eqnref{xi_B}, \eqnref{phi'0} and \eqnref{b'i5}. Note that we may
always add any multiple of $\kh_j$ to
$A_{ij}$ and $E_j$ without altering \Eqsref{aij,ci} and
\eqnref{ej,f}. In order to get unambiguous expressions, we require
that $A_{ij}\kh^j=E_j\kh^j= 0$. This yields that
\beqa
A_{ij} & = & \frac{1}{\phiinf} \sqrt{\frac{E}{2 \pi r^3 \sin^3
    \theta}} i e^{i \frac{\pi}{4}} \Big(- \sin^2 \theta \de_{ij}
+\khpd_i \khpd_j + \kh_i
\kh_j - \frac{\cos \varphi - \cos^2 \theta}{\sin^2 \theta}\khpd_i
\kh_j \Big) \qquad \\
C_i & = & \frac{1}{\phiinf} \sqrt{\frac{E}{2 \pi r^3 \sin^3 \theta}} i e^{i
   \frac{\pi}{4}} \sin \theta \Big( \kh_i - \frac{\cos \varphi -
\cos^2 \theta}{\sin^2 \theta} \khpd_i \Big) \\
E_i & = & \frac{1}{\phiinf} \sqrt{\frac{E}{2 \pi r^3 \sin^3 \theta}} i e^{i
   \frac{\pi}{4}} \sin \theta \Big(\khpd_i - \frac{\cos \varphi -
\cos^2 \theta}{\sin^2 \theta} \kh_i \Big) \\
F & = & \frac{1}{\phiinf} \sqrt{\frac{E}{2 \pi r^3 \sin^3 \theta}} i e^{i
   \frac{\pi}{4}} \left(\cos^2 \theta - \cos \varphi \right).
\eeqa
We see that the coefficients $C_i$ and $E_i$ are closely related, and
that the matrix $A_{ij}$ has a certain symmetry property. This means
that the roles of incoming and outgoing waves may be interchanged,
just as one would expect.

\vspace{2cm}
\noindent
\textbf{Acknowledgments:} M.H.~is a Research Fellow at the Royal
Swedish Academy of Sciences. We would like to thank Martin Cederwall
for discussions.

\clearpage
\bibliographystyle{utphysmod3b}
\bibliography{thomson}

\end{document}

%% file: macros.tex
\newcommand {\beq} {\begin{equation}}
\newcommand {\eeq} {\end{equation}}
\newcommand {\beqa} {\begin{eqnarray}}
\newcommand {\eeqa} {\end{eqnarray}}
\newcommand {\beqan} {\begin{eqnarray*}}
\newcommand {\eeqan} {\end{eqnarray*}}

\newcommand {\nn} {\nonumber}
\newcommand {\ph}[1]{\phantom{#1}}
\newcommand {\ie}{i.e.~}
\newcommand {\eg}{e.g.~}
\newcommand {\cf}{cf.~}

\newcommand {\sss} {\scriptscriptstyle}

\newcommand{\ordo}{\ensuremath{\mathcal{O}}}

\newcommand{\de}{\ensuremath{\delta}}

\newcommand{\eps}{\ensuremath{\epsilon}}
\newcommand{\veps}{\ensuremath{\varepsilon}}
\newcommand{\ka}{\ensuremath{\kappa}}
\newcommand{\la}{\ensuremath{\lambda}}

\newcommand{\La}{\ensuremath{\Lambda}}
\newcommand{\si}{\ensuremath{\sigma}}
\newcommand{\Si}{\ensuremath{\Sigma}}
\newcommand{\om}{\ensuremath{\omega}}
\newcommand{\Om}{\ensuremath{\Omega}}

\newcommand{\desi}{\ensuremath{\delta_{\Si}}}
\newcommand{\Omsi}{\ensuremath{\Omega_{\Si}}}

\newcommand{\volsi}{\ensuremath{\mathrm{Vol}_{\Si}}}

\newcommand{\mathH}{\mathaccent"707D}
\newcommand{\mathHb}[1]{{\mathop{\kern0pt#1}\limits^{\,\sss
      \prime\prime}\vphantom{#1}}}
\newcommand{\Xdot}{\ensuremath{\dot{X}}}
\newcommand{\Xddot}{\ensuremath{\ddot{X}}}
\newcommand{\Xprim}{\ensuremath{\acute{X}}}
\newcommand{\Xdprim}{\ensuremath{\mathH{X}}}
\newcommand{\bx}{\ensuremath{\mathbf{x}}}

\newcommand{\kh}{\ensuremath{\hat{k}}}

\newcommand{\khpd}{\ensuremath{{\hat{k}'}}}

\newcommand{\Hst}{\ensuremath{H_{\mathrm{st}}}}

\newcommand{\Hin}{\ensuremath{H_{\mathrm{in}}}}
\newcommand{\Hinu}{\ensuremath{H^{\mathrm{in}}}}
\newcommand{\Hel}{\ensuremath{H_{\mathrm{el}}}}

\newcommand{\Hout}{\ensuremath{H_{\mathrm{out}}}}
\newcommand{\Houtu}{\ensuremath{H^{\mathrm{out}}}}
\newcommand{\Bin}{\ensuremath{B_{\mathrm{in}}}}
\newcommand{\Binu}{\ensuremath{B^{\mathrm{in}}}}
\newcommand{\Bel}{\ensuremath{B_{\mathrm{el}}}}
\newcommand{\Belu}{\ensuremath{B^{\mathrm{el}}}}
\newcommand{\Bout}{\ensuremath{B_{\mathrm{out}}}}
\newcommand{\Boutu}{\ensuremath{B^{\mathrm{out}}}}
\newcommand{\phist}{\ensuremath{\phi_{\mathrm{st}}}}

\newcommand{\phiin}{\ensuremath{\phi_{\mathrm{in}}}}
\newcommand{\phiout}{\ensuremath{\phi_{\mathrm{out}}}}

\newcommand{\Xth}{\ensuremath{X_{\mathrm{Th}}}}
\newcommand{\Xthu}{\ensuremath{X^{\mathrm{Th}}}}
\newcommand{\Xdotth}{\ensuremath{\Xdot_{\mathrm{Th}}}}
\newcommand{\Xdotthu}{\ensuremath{\Xdot^{\mathrm{Th}}}}
\newcommand{\Xprimth}{\ensuremath{\Xprim_{\mathrm{Th}}}}
\newcommand{\Xprimthu}{\ensuremath{\Xprim^{\mathrm{Th}}}}
\newcommand{\Xddotth}{\ensuremath{\Xddot_{\mathrm{Th}}}}
\newcommand{\Xddotthu}{\ensuremath{\Xddot^{\mathrm{Th}}}}
\newcommand{\Xdprimth}{\ensuremath{\Xdprim_{\mathrm{Th}}}}
\newcommand{\Xdprimthu}{\ensuremath{\Xdprim^{\mathrm{Th}}}}

\newcommand{\dest}{\ensuremath{\de_{\mathrm{st}}}}
\newcommand{\deth}{\ensuremath{\de_{\mathrm{Th}}}}

\newcommand{\Omst}{\ensuremath{\Om_{\mathrm{st}}}}
\newcommand{\Omth}{\ensuremath{\Om_{\mathrm{Th}}}}
\newcommand{\phiinf}{\ensuremath{\phi_{\infty}}}
\newcommand{\vepphi}{\ensuremath{\frac{\veps^2}{\phiinf}}}
\newcommand{\vepphis}{\ensuremath{\frac{\veps^4}{\phiinf^2}}}

\newcommand {\we} {\wedge}
\newcommand {\pa} {\partial}

\newcommand {\pamu} {\partial^{\mu}}
\newcommand {\pamd} {\partial_{\mu}}
\newcommand {\panu} {\partial^{\nu}}
\newcommand {\pand} {\partial_{\nu}}
\newcommand {\paru} {\partial^{\rho}}
\newcommand {\pard} {\partial_{\rho}}

\newcommand {\dxm} {dx^{\mu}}
\newcommand {\dxmn} {\dxm \we dx^{\nu}}
\newcommand {\dxmnr} {\dxmn \we dx^{\rho}}

\newcommand{\eqnlab}[1]{\label{eqn:#1}}
\newcommand{\figlab}[1]{\label{fig:#1}}

\newcommand{\eqnref}[1]{(\ref{eqn:#1})}

\newcommand{\Eqnref}[1]{Eq.~(\ref{eqn:#1})}
\newcommand{\Figref}[1]{Fig.~\ref{fig:#1}}

\newcommand{\Eqsref}[1]{Eqs.~(\ref{eqn:#1})}

\newcommand{\seclab}[1]{\label{sec:#1}}

\newcommand{\secref}[1]{\ref{sec:#1}}